\documentclass[aps,prb,reprint]{revtex4-2}
\usepackage{graphicx}
\usepackage{amsmath}
\usepackage{amssymb}
\usepackage{gensymb}
\usepackage{hyperref}
\usepackage{bm}
\usepackage{color}

\begin{document}
\title{Noncollinear ferrielectricity and hydrogen-induced ferromagnetic polar half-metallicity in MnO$_3$Cl}
\author{Xinyu Yang}
\author{Jun Chen}
\author{Shan-Shan Wang}
\email{wangss@seu.edu.cn}
\author{Shuai Dong}
\email{sdong@seu.edu.cn}
\affiliation{Key Laboratory of Quantum Materials and Devices of Ministry of Education, School of Physics, Southeast University, Nanjing 211189, China}

\begin{abstract}
Collinear dipole orders such as ferroelectricity and antiferroelectricity have developed rapidly in last decades. While, the noncollinear dipole orders are rarely touched in solids. Noncollinear dipole orders can provide a route to realize ferrielectricity. Based on first-principles calculations, an inorganic molecular crystal MnO$_3$Cl has been demonstrated to own intrinsic noncollinear ferrielectricity, which originates from the stereo orientations of polar molecules. The large negative piezoelectricity effect ($d_{33} \sim$ -27 pC/N) is also predicted. A strong light absorption and moderate optical anisotropy are found for this molecular crystal in the ultraviolet light window. Additionally, by electron doping via hydrogen intercalation, a ferromagnetic polar half-metals can be obtained. Our study here provide a material platform to explore the intriguing physics of noncollinear ferrielectricity and potential applications in devices.
\end{abstract}
\maketitle
	
\section{Introduction}
Ferroic materials, possessing switchable vectors like polarization and magnetization, are one of favorable topics in condensed matter community and of potential applications in memory devices~\cite{schmid1994, Ramesh2019, Khomskii2009, SCOTT2007, RevModPhys.78.1185, Dong2020}. In the framework of Laudau-Ginzburg theory, the magnetic and electrical dipole orders have the one-to-one correspondence. In magnets, both collinear and noncollinear magnetic orders have been extensively investigated~\cite{Ramirez1994, Mostovoy2007, Kimura2007}. Many progress have been reported towards the noncollinear magnetism, which can give rise to various physical properties, inculding the anomalous Hall effect, skyrmions and magento-optical Kerr effects~\cite{Tokura, RevModPhys.82.1539, PhysRevB.92.144426}. 
However, noncollinear polarizations remain in early stage. In addition, the noncollinear dipole texture contains the ferroelectric (FE) and antiferroelectric (AFE) modes, which can lead to ferrielectricity. Till now,  only a few works have focused on it. It is found that the noncollinear dipole order has been realized in perovskite deviants in experiments~\cite{Yadav2016, science.aay7356}. Theoretical study predicts that strained BiFeO$_{3}$ can exhibit noncollinear ferrielectric polarization \cite{PhysRevLett.109.057602, PhysRevLett.112.057202, PhysRevLett.107.117602}. Very recently, a two-dimensional dioxydihalides family is also predicted to display intrinsic noncollinear ferrielectricity and thus generates topological domains and negative piezoelectricity, but yet to be verified by experiments~\cite{PhysRevLett.123.067601}. The lack of material platforms make it hard to uncover the underlying physical mechanism of noncollinear electrical dipole orders. In this scenario, it is extremely necessary to broaden the types of noncollinear polar materials to thoroughly study the unexplored properties.

On the other hand, molecular ferroelectricity has undergone a rapid development and attracted extensive attentions, due to its structural flexibility, bio-compatibility and lightweight~\cite{science.aai8535,Xiong2020,XiongrengenD3CS00262D,liu2022jpcl, Han2019nc}. Many efforts have been devoted to the organic molecular ferroelectricity but the noncollinearity mostly comes from the stereochemical structure of bulky organic groups. Then the switching of canting dipole would be relative slow and the switching path would be complex. Compared with the organic molecules, the pure inorganic molecules own very small molecular unit, which may provide the feasibility of fast switching comparable to widely used inorganic ferroelectric perovskites. Such noncollinear dipoles in pure inorganic molecules are not common.
	
Recectly, an inorganic molecular bulk with high valence manganese (Mn$^{7+}$) atoms, MnO$_3$Cl, was successfully synthesized in the experiment ~\cite{Dr2006zaac}. The MnO$_3$Cl is made up of tetrahedral polar molecular units with weak inter-molecular interactions, which may provide an ideal material platform to explore noncollinear dipole orders.

In this work, based on first-principles calculations and Monte-Carlo simulations, we demonstrated that MnO$_3$Cl displays a new type of noncollinear ferrielectric orders. A special switchable path of electric polarization through molecular rotation-flip model has been constructed. Besides, the intriguing negative piezoelectric effect and optical properties are also revealed in this molecular crystal. Upon intercalating hydrogen atoms to its interstitial positions, there is a distinct transition from nonmagnetic ferrielectricity to ferromagnetic polar half-metallicity, where $100\% $ spin polarization and polarity coexist in the single crystal Mn(OH)$_3$Cl.

\section{Computational methods}
First-principles calculations based on density functional theory (DFT) are performed with the projector augmented-wave (PAW) pseudopotentials as implemented in the Vienna {\it ab initio} Simulation Package (VASP) ~\cite{kresse1996Prb}. The exchange-correlation functional is treated using Perdew-Burke-Ernzerhof (PBE) parametrization of the generalized gradient approximation (GGA) ~\cite{perdew1996Prl}. More tests with different exchange-correlation functionals can be found in Table 1. The energy cutoff is fixed to $500$ eV. The $\Gamma$-centered $4\times3\times6$ Monkhorst-Pack \textit{k}-mesh is adopted. The vdW correction DFT-D3 method is applied \cite{Stefan2010JCP}. The convergence criterion for the energy is $10^{-6}$ eV for self-consistent iteration, and the Hellman-Feynman force is set to $0.01$ eV/\AA{} during the structural optimization. Ferrielectric polarization is calculated using the standard Berry phase method ~\cite{King1993prb}.

The {\it exciting} package is adopted to calculate the nonlinear optical susceptibility tensors for second harmonic generation (SHG) of MnO$_3$Cl~\cite{Gulans2014JPCM}. During the SHG calculation, the tolerence factor is set as $1\times10^{-3}$ to avoid singularities.

The optical parameters are performed using the frequency dependent dielectric functions expressed as $\epsilon(\omega)=\epsilon_1(\omega)+i\epsilon_2(\omega)$. And the optical absorption coefficient $\alpha(\omega)$ has a relationship with the real part of the dielectric function $\epsilon_1(\omega)$ and the imaginary component $\epsilon_2(\omega)$:
\begin{equation}
	\alpha(\omega)=\frac{2\sqrt{2}\pi E}{hc}[\sqrt{\epsilon_1^2(\omega)+\epsilon_2^2(\omega)}-\epsilon_1(\omega)]^\frac{1}{2},
\end{equation}

where $E$ is the incident photon energy, $h$ is the Planck constant and $c$ is the speed of light in vacuum. $\epsilon_1(\omega)$ and $\epsilon_2(\omega)$ can also be used to calculate the refactive index $n(\omega)$ and the extinction cofficient $k(\omega)$ using the following equations~\cite{SUJITH2023JPCS}: 
\begin{equation}
	n(\omega)=\frac{1}{\sqrt{2}}[\sqrt{\epsilon_1^2(\omega)+\epsilon_2^2(\omega)}+\epsilon_1(\omega)]^\frac{1}{2},
\end{equation}

\begin{equation}
	k(\omega)=\frac{1}{\sqrt{2}}[\sqrt{\epsilon_1^2(\omega)+\epsilon_2^2(\omega)}-\epsilon_1(\omega)]^\frac{1}{2},
\end{equation}

Considering that the PBE function usually underestimates the band gap of the semiconductor, we carry out a PBE-level optical properties calculation with a corrected parameter $\Delta E_{g,\rm{HSE}}$, the energy diﬀerence between the PBE band gap and Heyd-Scuseria-Ernzerhof hybrid functional band gap~\cite{Huang2015PhysRevB}.

For the Mn(OH)$_3$Cl, the Hubbard $U$ is applied using the Dudarev parametrization ~\cite{dudarev1998Prb}. As reported previously, a correction of $U_{\rm eff}=3$ eV is imposed on Mn's $3d$ orbitals ~\cite{wangprb2023}. In addition, the {\it ab initio} molecular dynamics (AIMD) simulation in the NVT ensemble lasts for 6 ps with a time step of 2 fs. The Nose-Hoover method is used for temperature control.~\cite{Martyna1992JCP}. Moreover, the Monte Carlo (MC) simulations based on Heisenberg model are applied to estimate the magnetic transition temperatures. A 18 × 18 × 18 lattice with periodic boundary condition is adopted in MC simulations. The initial $5 \times 10^4$ MC steps (MCSs) are performed for thermal equilibrium, then another $5 \times 10^4$ MCSs are reserved as statistical average.

\section{Results and discussion}
\subsection{Noncollinear ferrielectricity} 

The MnO$_3$Cl single crystal consists of unique zero dimensional ($0$D) structures, which extends into  three dimensional ($3$D) molecular frameworks through the weak vdW interactions.
Each MnO$_3$Cl owns a dipole, pointing along the Mn-Cl bonding direction (mainly along the $b$-axis). The dipole-dipole interactions prefer: 1) the antiparallel alignment between side-by-side neighbors (A $\&$ B) along the $c$-axis; 2) the parallel alignment between head-to-tail neighbors (A $\&$ A’) along the $b$-axis, as displayed in Fig. S1 of the Supplemental Material (SM) \cite{sm}. 

\begin{figure}
	\includegraphics[width=0.48\textwidth]{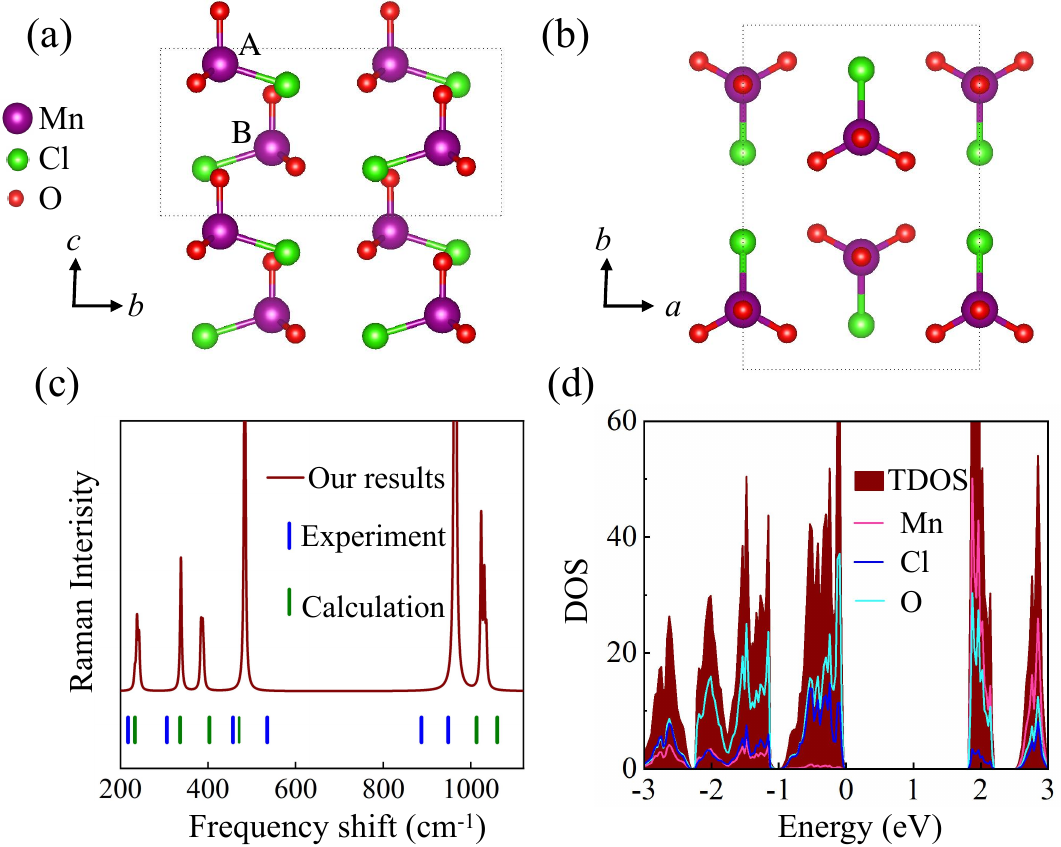}
	\caption{(a-b) The side and top views of MnO$_3$Cl crystal. The primitive cell is indicated by the black dotted rectangula. (c) The DFT calculated raman spectra of MnO$_3$Cl crystal. The characteristic peak points of previous reported values (experiment $\&$ calculation) are shown as bars for comparison. (d) The projected density of states (pDOS) structure of MnO$_3$Cl crystal.}
	\label{F1}
\end{figure}	

However, this collinear dipole order is not the most stable one. As shown in Fig.~S1 \cite{sm}, the antiphase canting of the upper and lower molecules (A $\&$ B) can reduce the energy for 0.17 eV/f.u. (at 17.4°). Such energy reduction can be intuitively attributed to the Coulomb repulsion. After canting, the nearest O1-O9 distance between A $\&$ B is enlarged, while the nearest Cl2-O9 distance between A $\&$ B is shorten. By inputting the concrete values, a rough estimation leads to a positive energy gain from such molecular canting. The stable structure within a unit cell is depicted in Figs.~\ref{F1}(a-b). It has a polar space group of $Cmc2_1$ with mirror and two-fold rotation symmetries. According to our calculation, the optimized lattice constant is in agreement with the experiment (see Table~1)~\cite{Dr2006zaac}. Furthermore, the DFT calculated Raman spectra of MnO$_3$Cl crystal is also consistent with previous reported values, as shown in Fig.~\ref{F1}(c). Figure 1(d) displays its projected density of states structure, indicating a semiconductor state with band gap of $\sim$1.83 eV. It is noted that electronic localization shows peak effects, and the peak effect at the Fermi level is mainly contributed by the O-$p$ orbital, suggesting its potential for efficient light absorption (to be discussed later).

\begin{table}
	\caption{The structure parameters of MnO$_3$Cl. The tested exchange-correlation functionals are GGA-PBE and Perdew-Burke-Ernzerhof-revised (PBEsol) of GGA. The experimental results (Expt.) are listed for comparison. The $a$, $b$ and $c$ are in units of \AA{}. Notably, the GGA-PBE correction makes lattice constants closer to the experimental values.}
	\begin{tabular*}{0.48\textwidth}{@{\extracolsep{\fill}}lccccc}
		\hline \hline
		&  $a$ & $b$ & $c$\\
		\hline
		Expt.~\cite{Dr2006zaac}      & 7.155 & 10.084  & 5.009 \\		
		GGA-PBE    & 6.985 & 10.166  & 5.000 \\
		GGA-PBEsol & 6.626 & 9.909   & 4.782 \\
		\hline \hline
	\end{tabular*}
	\label{Table 1}
\end{table}	

\begin{figure}
	\includegraphics[width=0.42\textwidth]{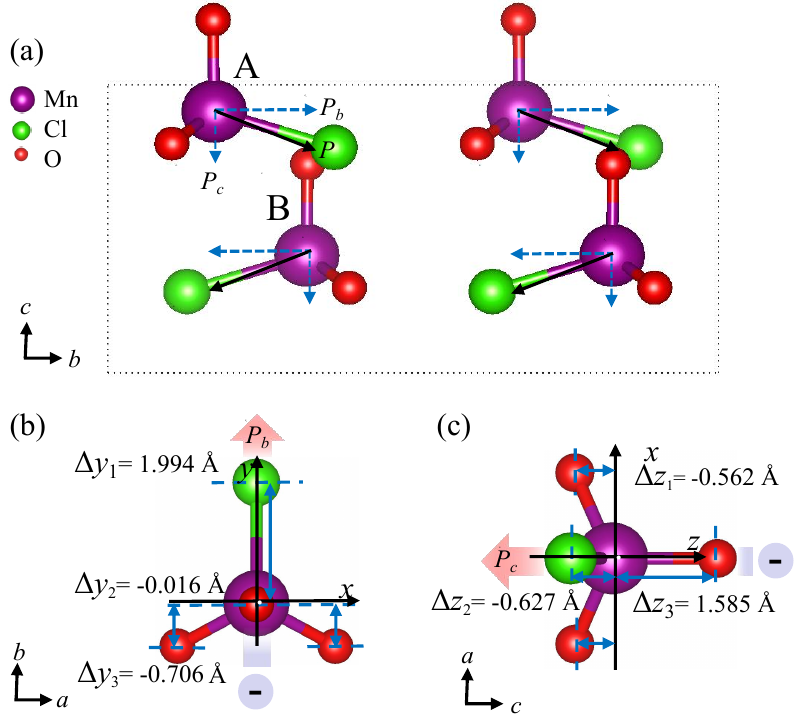}
	\caption{(a) Schematic of ferrielectric polarization in MnO$_3$Cl crystal. The $b$-components of molecules A and B are compensated, while the $c$-components are parallel. Taking the molecule A for example: (b) View from the $c$-axis. The $b$-axis projected distances between O$^{2-}$/Cl$^{-}$ and Mn$^{7+}$ are denoted as $\Delta y_1$,  $\Delta y_2$,  $\Delta y_3$. Then $d_b$=-$\Delta y_1$-2$\Delta y_2$-4$\Delta y_3$=0.862 $|e|$\AA{}, implying a net dipole along the $b$ direction. (c) View from the $b$-axis. The $c$-axis projected distances between O$^{2-}$/Cl$^{-}$ and Mn$^{7+}$ are denoted as $\Delta z_1$,  $\Delta z_2$,  $\Delta z_3$. Then $d_c$=-4$\Delta z_1$-$\Delta z_2$-2$\Delta z_3$=-0.295 $|e|$\AA{}, implying a net dipole along the $-c$ direction.}
	\label{F2}
\end{figure}

Each MnO$_3$Cl molecule contributes a dipole ($d$), as shown in Fig.~\ref{F2}, which can be estimated as:
\begin{equation}
\textbf{d}_{ij}=\sum_{j}\Delta \textbf{r}_{ij}q_{j}
\end{equation}
where $q_{j}$ is the charge of O$^{2-}$/Cl$^{-}$ ions, and $\Delta \textbf r_{ij}$ represents the offset vector of O$^{2-}$/Cl$^{-}$ relative to Mn$^{7+}$. Then the value of $\textbf{d}$ is estimated as (0, 0.862, -0.295) $|e|$\AA{} for molecule A, while it is (0, -0.862, -0.295) $|e|$\AA{} for molecule B.

\begin{figure}
	\includegraphics[width=0.5\textwidth]{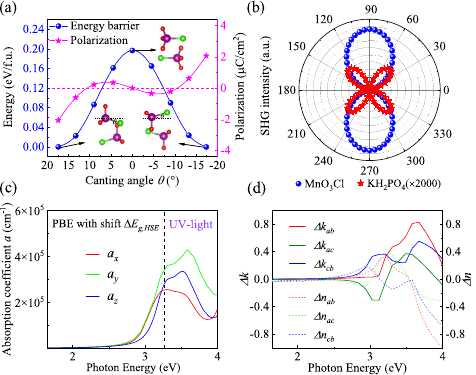}
	\caption{(a) {The energy barrier and polarization of possible ferrielectric switching path as a function of the canting angle $\theta$ of Mn-Cl bond from the $b$-axis for MnO$_3$Cl crystal.} (b) The comparison of calculated SHG angular plots of MnO$_{3}$Cl and KH$_{2}$PO$_{4}$ crystals on the $bc$ plane with $1064$ nm light. (c) Optical absorption spectra. Solid lines of different color represent the calculated results for MnO$_{3}$Cl (red and bule lines for polarization of light in the $ac$ plane and green line for polarization of light out-of-plane along the $b$-axis) Visible region:1.65-3.26 eV~\cite{Huang2015PhysRevB}. (d) Anisotropic optical properties of MnO$_{3}$Cl crystal. Birefringence: $\Delta n_{ij}=n_{j}-n_{i}$; Dichroism: $\Delta k_{ij}=k_{j}-k_{i}$. Here, $i$ and $j$ are any two orthogonal crystal axes: $a, b,$ and $c$.}
	\label{F3}
\end{figure}

Among the whole bulk structure, it displays a surprising noncollinear ferrielectric texture with the partially compensation of dipoles, namely the $b$-components of dipoles are compenstated, but the $c$-components of dipoles lead to a net switchable polarizations. Therefore, it can exhibit both ferroelectric ($P_c$: ferroelectric order parameter) and antiferroelectric ($P_b$: antiferroelectric order parameter) behaviors. These two components are coupled via the geometric relationship: $P_b^2$+$P_c^2$=$P^2$, where $P$ is almost a constant of MnO$_3$Cl molecule. 

Besides the intuitive estimation based on the point charge mode, we also calculated the rigorous polarization using the Berry phase method as implemented in VASP, which leads to $2.1$ $\mu$C/cm$^2$ along the opposite direction of the $c$-axis in the MnO$_3$Cl crystal. Thus the direction of dipole is consistent between the point charge model and Berry phase method. The value of polarization is comparable with that of CuInP$_{2}$S$_{6}$ ($ 2.55$ $\mu$C/cm$^2$) ~\cite {PhysRevB.56.10860}. 

The switching of polarization is another important properties for ferroic materials. As shown in Fig.~\ref{F3}(a), the ferrielectric switching process is also estimated using climbing image nudged elastic band (CINEB) method~\cite{Graeme2000JCP}. Assuming an electric field is applied along the polarization direction, polar molecules may rotate around the $a$-axis, leading to the net polarization flipping. Apart from alterations in bond angles, bond lengths also undergo corresponding changes (see Fig.~S2) \cite{sm}. The energy barrier of transformation proceeds through an antiferroelectric configuration (space group $P2_1/m$). And the switching energy barrier between the ferrielectric state and antiferroelectric state is $\sim0.2$ eV/f.u., suggesting the ferrielectric stability and switchability. 

In practice, the optical second-harmonic generation (SHG) is a sensitive and nondestructive tool to characterize non-centrosymmetric materials~\cite{S.prl2003, Abdelwahab2022NP}, in particular vital for ferrielectric materials. The lack of spatially inversion symmetry in MnO$_3$Cl crystals allows the intrinsic SHG signal. Here, the nonlinear optical susceptibility tensors ($d_{ij}$) can be expressed as a $3\times6$ matrix~\cite{Sutherland2003handbook}:

\begin{equation}
	d={\left[\begin{array}{cccccc}
			0&0&0&0&d_{15}&0\\
			0&0&0&d_{24}&0&0\\
			d_{31}&d_{32}&d_{33}&0&0&0
		\end{array}
		\right]}.
	\label{1}
\end{equation}

Then, the second harmonic polarization $P^{2\omega}$ can be expressed as \cite{yang2024jcp}:
\begin{equation}
	\begin{aligned}
	{\left[\begin{array}{c}
			P^{2\omega}_{a}\\
			P^{2\omega}_{b}\\
			P^{2\omega}_{c}
		\end{array}
		\right]}
	=&2\times {\left[\begin{array}{cccccc}
			0&0&0&0&d_{15}&0\\
			0&0&0&d_{24}&0&0\\
			d_{31}&d_{32}&d_{33}&0&0&0
		\end{array}
		\right]} 
	{\left[\begin{array}{c}
			E^2_{a}\\
			E^2_{b}\\
			E^2_{c}\\
			2E_{b}E_{c}\\
			2E_{a}E_{c}\\
			2E_{a}E_{b}\\
		\end{array}
		\right]}\\
	=&2\times {\left[\begin{array}{c}
		2d_{15}E_{a}E_{c}\\
		2d_{24}E_{b}E_{c}\\
		d_{31}E^2_{a}+d_{32}E^2_{b}+d_{33}E^2_{c}\\
	\end{array}
	\right]},
	\label{1}
\end{aligned}
\end{equation}	
where $\omega$ is the wave of frequency. $a/b/c$ are perpendicular crystalline axes here. The SHG intensity $I$ can be estimated as~\cite{Sutherland2003handbook}: 
\begin{equation}
	\begin{aligned}
		I \propto (P^{2\omega})^2=(P^{2\omega}_{a})^2+(P^{2\omega}_{b})^2+(P^{2\omega}_{c})^2	
	\end{aligned}
\end{equation}

For MnO$_3$Cl under perpendicular incident light on the $bc$ plane, polarization vector can be expressed as $\bm{E}=(E_a, E_b, E_c)=E(0,\cos\phi, \sin\phi)$. The SHG intensity can be expressed as: $I\propto(2d_{24}\sin2\phi)^2+(2d_{32}\cos^2\phi+2d_{33}\sin^2\phi)^2$, where $\phi$ is the angle between the crystalline $b$-axis and the electric field direction of incident light. Similarly, the SHG intensity of KH$_2$PO$_4$ on its $bc$ plane can be written as: $I\propto(2d_{14}\sin2\phi)^2$.

The calculated SHG tensors $d_{ij}$ of MnO$_3$Cl bulk at a wavelength of 1064 nm (which is frequently used in SHG experiments) are shown in Fig.~S3 of SM\cite{sm}, in comparison with a frequently-used reference KH$_{2}$PO$_{4}$~\cite{Eckardt1990IEEE}. Furthermore, the SHG angular plot of MnO$_3$Cl is calculated using these tensors, as shown in Fig.~\ref{F3}(b), it presents a bipolar behavior and the strongest signal appears when the angle $\phi$ reaches $90$° (the polarization direction of the incident light is along the $c$-axis), which is much stronger than that of KH$_{2}$PO$_{4}$.

\subsection{Negative piezoelectricity and optical properties}

Negative piezoelectricity refers to the phenomenon in which the polarization increases when the lattice is shrinking along the polarization direction. This exotic physical property is important in the study of ferroelectric/ferrielectric materials, but is rarely observed in experiments~\cite{Katsouras2016NM,Lu2019SA}. To investigate the piezoelectricity in this noncollinear ferrielectricity, piezoelectric stress coefficients $e_{ik}$ are calculated using density functional perturbation theory (DFPT) method~\cite{Gonze1997prb}. There are five independent nonzero piezoelectric stress coefficients $e_{ik}$ in the piezoelectric tensor matrix for point group $mm2$~\cite{deJong2015SD}:
\begin{equation}
	e={\left[\begin{array}{cccccc}
			0&0&0&0&0.12&0\\
			0&0&0&0.06&0&0\\
			0.09&0.02&-0.28&0&0&0
		\end{array}
		\right]}.
	\label{1}
\end{equation}

The calculated $e_{33}$ is -0.28 C/m$^2$. There is a stronger dipole moment within a smaller volume, due to the joint contribution of the noncollinear dipole orders and the intermolecular effects between the $0$D polar molecules. In addition, the elastic stiffness tensors ($C_{kj}$) are calculated by the energy-strain method using VASPKIT~\cite{WANG2021CPC}, which can be found in the SM~\cite{sm}. Then the piezoelectric strain coefficients $d_{ij}$ can be determined as follows:
\begin{equation}
	\begin{aligned}
		d_{ij} = \sum_{k=1}^{6}e_{ik}C_{kj}^{-1}
	\end{aligned}
\end{equation}

The longitudinal strain coefficient $d_{33}$ is -27.37 pC/N, which is 19 times that of ZrI$_2$ (-1.445 pC/N)~\cite{Ding2021PRM}, indicating a large negative piezoelectricity. 

Another noteworthy characteristic is the optical absorption of MnO$_3$Cl, as previously mentioned. The orthorhombic crystal MnO$_3$Cl has three principal dielectric constants: $\varepsilon_{xx}\neq\varepsilon_{yy}\neq\varepsilon_{xx}$. We label them as $\varepsilon_{a}$, $\varepsilon_{b}$ and $\varepsilon_{c}$ respectively, as calculated in Fig.~S4 of the SM~\cite{sm}. Figure~\ref{F3}(c) shows the calculated optical absorption cofficient $\alpha$ as a function of photon energy in MnO$_3$Cl. The first peak of optical absorption spectum can be seen in the energy windows of ultraviolet light (UV-light) and the absorption coefficient of light polarization along the $b$-axis reaches a large magnitude ($\sim$ $10^5$ cm$^{-1}$),  implying its strong UV-light absorption. Additionally, birefringence and dichroism can be utilized to assess optical anisotropy of the optical properties, quantifying the difference between any two components of $ n$ and $k$ along the principal crystal axes.
Figure 3(d) shows anisotropy exists along the direction $b$, as reflected by the large positive values ($\Delta k_{ab}, \Delta k_{cb}$) in the energy range of 3.26-4.0 eV and high values of $\Delta n_{ab}$ and $\Delta n_{cb}$.

\subsection{Ferromagnetic polar half-metallicity }
In general, the screening effect of free carriers on dipole-dipole interaction makes it difficult for coexistence of metallic properties and polarization. Moreover, it is even more challenging to form ferromagnetic polar half-metals with appropriate electronic structures~\cite{Puggioni2018prm,Zhang2024NM}.
 
MnO$_3$Cl crystal with Mn's $3d^0$ configuration behaves non-magnetic. Even though, the variable valence of Mn provides the feasibility to induce its magnetism via electron doping. As mentioned before, $3$D MnO$_3$Cl is spatially loose due to the vdW interactions, which possesses adequate interstitial positions for intercalating hydrogen atoms~\cite{Gong2021prb, you2022mtp, zhang2022ami}. 

Considering all possible H intercalation positions of MnO$_3$Cl and different stacking modes ($\alpha$-state vs $\beta$-state), we optimized all the possible structures. Mn(OH)$_3$Cl undergoes a structural reconstruction during the relaxation. H-intercalation causes two molecules with tetrahedral configurations to combine into a conventional octahedron, as indicated in Fig.~S5 of the SM~\cite{sm}. According to our DFT calculation, the configuration $\beta$ owns the lower energy. The AIMD simulation at 200 K further conﬁrms its thermal stability after the structural reconstruction (see Fig.~S6 in the SM)~\cite{sm}. Besides, we have also estimated available equilibrium chemical potential region to be H-rich/Mn-poor for stable growth of Mn(OH)$_3$Cl in Fig.~S7 (see SM for details)~\cite{sm}.

To determine Mn(OH)$_3$Cl's magnetic ground state, four most possible magnetic orders are compared, as shown in Fig.~S8 of the SM~\cite{sm}. According to our calculation, the energy of FM is the lowest one, implying a FM ground state (see Table 2). 

\begin{table}
	\caption{Calculated energies (in units of meV/f.u.) of different stacking models ($\alpha$-state vs $\beta$-state) with four possible magnetic orders in Mn(OH)$_{3}$Cl crystal .  The $\beta$-state with FM order is taken as the reference, which has the lowest energy.}
	\begin{tabular*}{0.48\textwidth}{@{\extracolsep{\fill}}lcccc}
		\hline \hline
		&AFM 1&AFM 2&AFM 3&FM\\
		\hline
		$\alpha$-state &207.02&183.55&204.39&178.29\\
		$\beta$-state &25.29&2.04&24.85&0\\
		\hline\hline
	\end{tabular*}
	\label{Table 2}
\end{table}

\begin{figure}
	\includegraphics[width=0.5\textwidth]{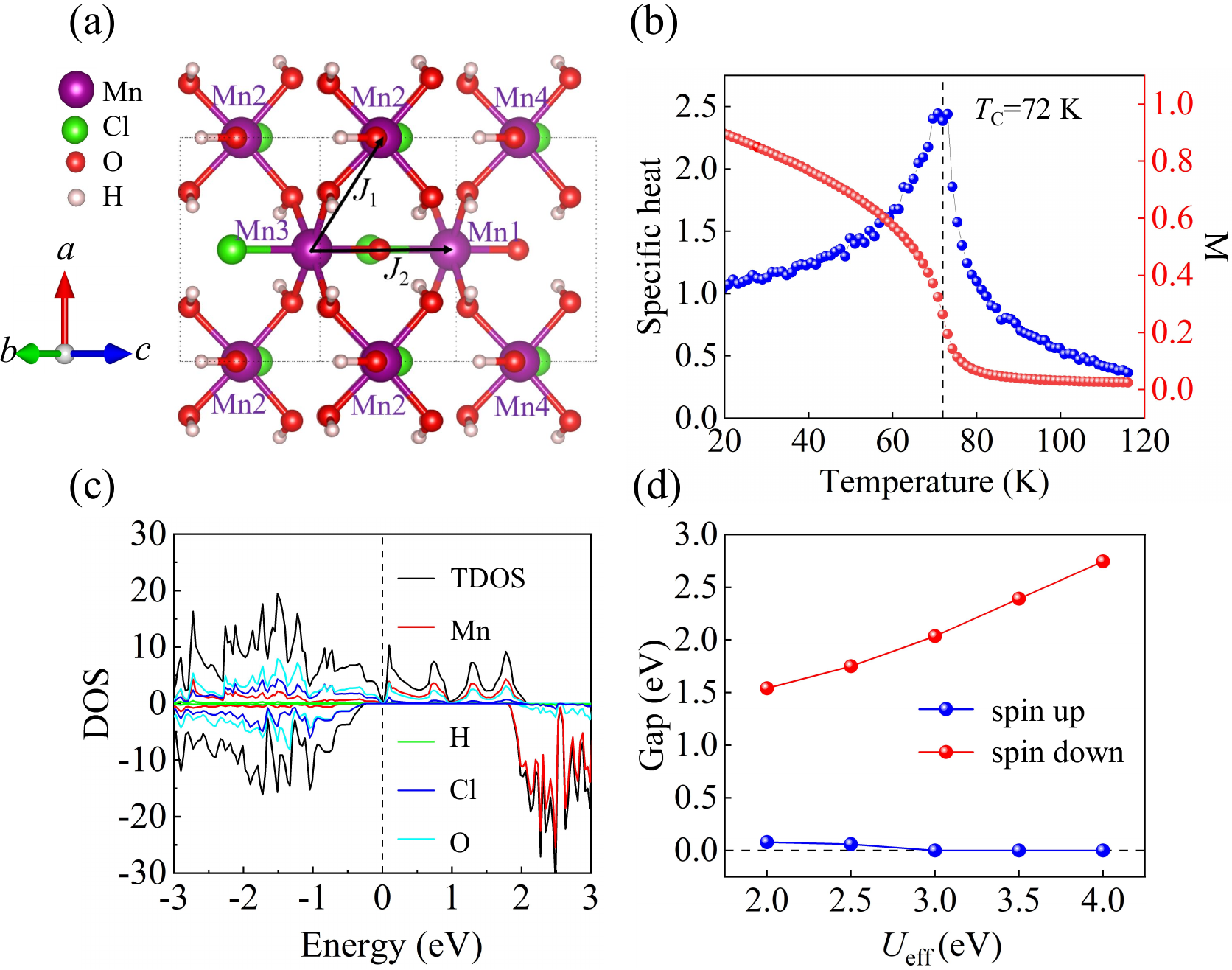}
	\caption{(a) Exchange $J'$s are indicated in Mn(OH)$_3$Cl. (b) MC calculated specific heat and normalized ferromagnetic order parameter $M$ as a function of temperature. The peak of specific heat indicates the magnetic transition. (c) The pDOS structure of Mn(OH)$_3$Cl. (d) The spin-up and spin-down energy gaps as functions of various $U_{\rm eff}$ for Mn(OH)$_3$Cl. }
	\label{F4}
\end{figure}

Our calculation shows a total magnetization of $\sim3\mu_{\rm B}$/Mn, as expected for its $3d^3$ configuration, implying Mn$^{4+}$. Based on the optimized structure of FM ground state, the nearest- and next-nearest-neighbor exchange coupling parameters $J_1$ and $J_2$, are derived by mapping the DFT energy to the Heisenberg model:
\begin{equation}
	H=\sum_{<ij>}J_{ij}\vec{S}_i\cdot\vec{S}_j + \sum_{i}[A_x(\vec{S}_{i}^x)^2+A_z(\vec{S}_{i}^z)^2],
\end{equation}

where the first item is the effective exchange interaction. Since the third neighbor exchange parameters are small, only $J_1$ and $J_2$ are considered, as indicated in Fig.~\ref{F4}(a). $\vec{S}_i$ is normalized spin vector at site $i$; The last item is the magnetic anisotropy, and a positive $A_x$ ($A_y$) implies the magnetic hard axis. 

The coefficients $J_1$ and $J_2$ are estimated as -6.68 and -1.15 meV/f.u. (see calculation details in the SM~\cite{sm}). And $A_x$ ($A_y$) is 0.015 (0.113) meV/f.u.. According to the above DFT-derived coefficients, the specific heat and ferromagnetic order paramater $M$ are calculated using the MC method, indicating a Curie temperature $T\rm{_C}$ of 72 K, as shown in Fig.~\ref{F4}(b).

Besides, the pDOS suggests a half-metallic state, as depicted in Fig.~\ref{F4}(c). The spin-down channel shows a large energy gap of $\sim2$ eV, while the spin-up channel exhibits metallic state, which are mainly contributed by the partially empty Mn $d$ and O $p$ orbitals. Such a transport behavior makes it a promising candidate for spintronic applications.

In general, the electronic correlation may influence the electronic structures. Therefore, various values of effective Coulombic potential ($U_{\rm eff}$=2.0, 2.5, 3.0, 3.5, 4 eV) for Mn atoms have been applied, to observe the evolution of the energy gap (see Fig.~\ref{F4}(d). As the value of $U_{\rm eff}$ increases, the energy gap of the spin-down channel also expands, while the spin-up channel exhibits almost zero band gap across the entire energy range. This result suggests its robust half-metallic state. In addition, the evolution of spin-polarized band structures for Mn(OH)$_3$Cl as a function of $U_{\rm eff}$ has also been depicted in Fig.~S9 of the SM~\cite{sm}. Here, the symmetry of Mn(OH)$_3$Cl is reduced to monoclinic structure with a polar space group $Pm$. Therefore, the polarization along the $b$-axis and $c$-axis is nonzero, which originates from the intrinsic differences in the chemical environments around Cl and O atoms. The coexistence of polarity and ferromagnetic half-metallicity may offer significant multifunctional application in future.

\section{Conclusion}
In summary, based on first-principles calculations, we demonstrate that noncollinear ferrielectricity, large negative piezoelectricity and ultraviolet light absorption can be generated in MnO$_3$Cl crystal. Besides, we show that the intercalation of hydrogen atoms into the MnO$_3$Cl results in signiﬁcant structural reconstruction. Meanwhile, a phase transition from nonmagnetic ferrielectricity to ferromagnetic polar half-metallicity also occurs spontaneously accompanying this reconstruction. These results indicate that our work opens a promising avenue for future studies of noncollinear ferrielectric properties and potential design of multifunctional materials.

\begin{acknowledgments}
This work was supported by the Postgraduate Research $\And$ Practice Innovation Program of Jiangsu Province (Grant No. KYCX24$\underline{~}0361$) and National Natural Science Foundation of China (Grant No. 12325401 and 12104089). Most calculations were done on the Big Data Computing Center of Southeast University.

\end{acknowledgments}
	
\bibliography{reference}
\end{document}